\documentclass[aps,prl,amsmath,superscriptaddress,longbibliography]{revtex4-2}
\usepackage{graphicx}
\usepackage{amsmath}
\usepackage{amsfonts}
\usepackage{amssymb}
\usepackage{braket}
\usepackage{bm}
\usepackage{multirow}
\usepackage{color}
\usepackage[normalem]{ulem}
\usepackage{mathrsfs}
\usepackage{mathtools}
\usepackage{float}
\usepackage{booktabs}
\usepackage{amsmath}

\makeatletter

\newcommand{\Rmnum}[1]{\expandafter\@slowromancap\romannumeral #1@}
\makeatother

\usepackage[breaklinks]{hyperref}
\hypersetup{colorlinks=true, linkcolor=blue, citecolor=blue, filecolor=blue, urlcolor=blue}

\AtBeginDocument{%
    \newwrite\bibnotes
    \def\bibnotesext{Notes.bib}
    \immediate\openout\bibnotes=\jobname\bibnotesext
    \immediate\write\bibnotes{@CONTROL{REVTEX41Control}}
    \immediate\write\bibnotes{@CONTROL{%
    apsrev41Control,author="08",editor="1",pages="1",title="1",year="1"}}
     \if@filesw
     \immediate\write\@auxout{\string\citation{apsrev41Control}}%
    \fi
}%

\begin{document}

\title{Pressure-Induced Topological Dirac Semimetallic Phase in KCdP}

\author{Shivendra Kumar Gupta}
\email{shivendrakumarg900@gmail.com}
\affiliation{Department of Physics, Visvesvaraya National Institute of Technology, Nagpur, 440010, India \\}
\author{Nikhilesh Singh}
\author{Saurabh Kumar Sen}%
 \author{Nagarjuna Patra}%
\author{Poorva Singh}
\email{poorvasingh@phy.vnit.ac.in}
\affiliation{Department of Physics, Visvesvaraya National Institute of Technology, Nagpur, 440010, India \\}


\begin{abstract}
Dirac semimetals (DSMs), characterized by linear dispersion relations in their electronic band structure, have gained prominence due to their unique topological features and potential applications in electronic devices. Through systematic calculation, we explore the electronic structure evolution of KCdP under varying negative pressure conditions. Our findings reveal a compelling transition from a normal semiconductor to a triple point semimetal when spin-orbit coupling (SOC) is not introduced, whereas in the SOC case, it converts into a Dirac semimetallic state in KCdP under negative triaxial pressure. The electronic band structure exhibits distinct Dirac cones at the Fermi level, indicating the presence of massless Dirac fermions. Moreover, the negative pressure-induced Dirac semimetallic phase in this compound is found to be robust and is protected by crystal symmetry. We provide a symmetry analysis of the bandgap, Fermi surface, Fermi velocity, and other relevant electronic properties, offering insights into the pressure-driven phase transition in KCdP. The tunability of this material under external pressure suggests the potential utility in next-generation electronic devices and quantum technologies.

\end{abstract}

\maketitle

\section{Introduction}\label{sec1}

Topological materials showcase unique properties like dissipationless spin transport through topologically protected surface states \cite{ren2016topological}. The first topological material discovered was a topological insulator that showed dissipationless spin transport, along with the exchange of orbital band characteristics of conduction and valence bands with a small, finite bulk band gap \cite{betancourt2016complex, gupta2023coexistence, hasan2010colloquium, patel2025composite}. Later, similar kinds of surface properties were also found in semimetals exhibiting closing of the gap through bulk bands \cite{gao2019topological}. These kinds of semimetals generated much interest due to their unique properties and were mainly categorized into DSMs, Weyl semimetals, nodal line, and nodal surface semimetals \cite{bernevig2018recent}. DSMs have fourfold degenerate zero-dimensional nodal points having linear dispersion in band crossing distinct from Weyl semimetals, which possess twofold degenerate zero-dimensional nodal points, while nodal line or nodal surface semimetals have either four or twofold degenerate one-dimensional nodal line or ring or two-dimensional nodal plane \cite{huang2016topological}.

\begin{figure*}[t!]
\centering
\includegraphics[width=1\linewidth]{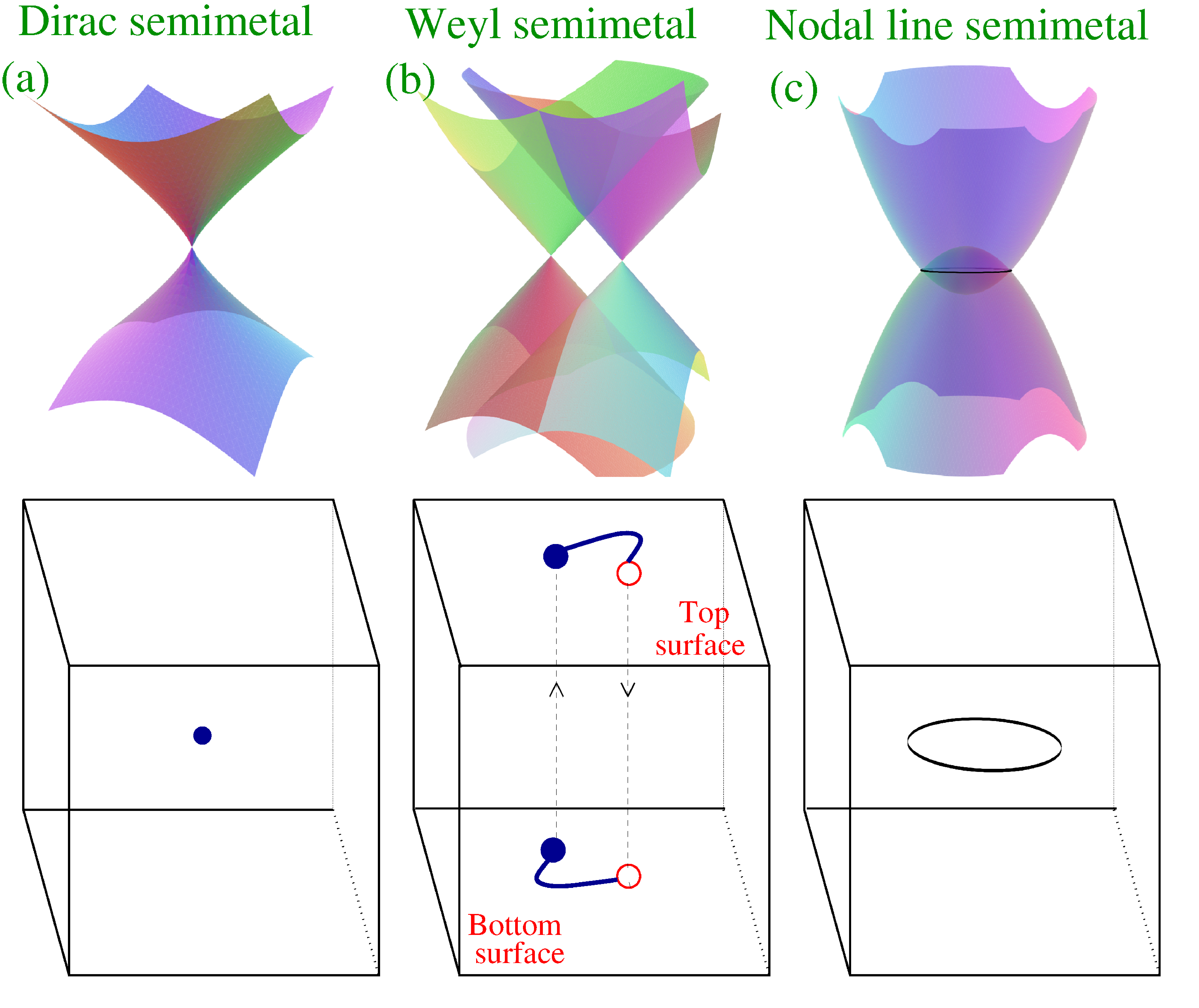}
\caption{\label{fig:epsart11} Schematic diagrams highlighting the distinct Fermi-surface features of (a) Dirac semimetals, (b) Weyl semimetals with surface Fermi arcs, and (c) nodal-line semimetals.}
\end{figure*}

For comparison, we explicitly contrast the electronic characteristics of DSMs with those of other topological semimetals in terms of their Fermi-surface topology. In a DSM, the conduction and valence bands cross at discrete points in momentum space, giving rise to point-like Fermi surfaces associated with Dirac nodes. In contrast, in Weyl semimetals, the Dirac point splits into pairs of Weyl nodes with opposite chirality, resulting in separated bulk Fermi pockets and the emergence of open Fermi-arc states on the surface Brillouin zone. Meanwhile, in nodal-line semimetals, band crossings occur along continuous lines or closed loops in momentum space, leading to extended loop-shaped Fermi surfaces accompanied by drumhead surface states in the surface electronic structure. Schematic diagrams illustrating the Fermi surfaces of different topological semimetals are shown in Figure-\ref{fig:epsart11}.

In the last 2 decades, the development of new topological materials and their classification has provided a path to understand their nature; however, there are some topological classes that need increased attention due to a lack of materials realization in the experimental framework. One of these types is topological DSM, a type of topological semimetal (TSM), which is identified by the crossing of conduction and valence bands at discrete points near the Fermi level in the Brillouin zone and linear dispersion relation  in all directions \cite{hou2021prediction}. Two Weyl points of opposite chirality overlap to form a 3D massless Dirac point, which is sensitive to perturbations. DSMs retain time reversal and inversion symmetries, showcasing unique properties such as giant diamagnetism, quantum magnetoresistance (MR) in the bulk, higher carrier mobility, oscillating quantum spin Hall effect, Fermi velocity near the Dirac cone, and the presence of Fermi arcs or Dirac points on the surface. Moreover, by breaking different symmetries, a topological Dirac semimetal (TDSM) can be converted into other exotic phases, such as topological insulators, Weyl semimetals, axion insulators, and topological superconductors \cite{liang2015ultrahigh,liu2014stable}. These  properties, along with the study of related topological quantum phase transitions makes DSMs ideal candidates to investigate.

 There are three prerequisites for a material to be a DSM. The first and foremost is the fourfold essential degeneracy enforced by nonsymmorphic symmetry at a high symmetry point in the Brillouin zone. $BiO_2$ and $BiZnSiO_4$ are such materials that theoretically fall under this category \cite{young2015dirac}, but are not experimentally verified yet. Secondly, the emergence of the accidental Dirac nodes at critical points of topological phase transition in special topological insulator materials having large Rashba splitting, Weyl semimetals, nodal line semimetals, and triple point semimetals. 3D Dirac fermions can be observed at a single quantum critical point, as they do not carry a Chern number; however, it requires extreme fine-tuning of alloy composition, which makes its experimental realization challenging \cite{yang2014classification}. There exists a third category as well: an accidental band-crossing DSM, induced by band inversion and protected by the appropriate axial rotational symmetry, which makes it robust over a finite range of Hamiltonian parameters. $Na_3 Bi$ \cite{liu2014discovery} and $Cd_3 As_2$ \cite{liang2015ultrahigh} are such materials that have been confirmed experimentally as topological DSMs under this category \cite{peng2018predicting}. Additionally, DSMs are classified into type I, type II, and type III, based upon the tilting of the bands resulting in the obeying or breaking of Lorentz invariance. Type I DSMs exhibit negative magnetoresistance across all directions while obeying Lorentz invariance, a feature found in high-energy physics. This DSM phase has been identified in a variety of crystal structures \cite{liu2014stable,young2012dirac,wang2013three,liu2014discovery, shende2023first}. Type II DSMs are characterized by the crossing of topologically protected and strongly tilted bands, creating electron and hole-like pockets along with the breaking of Lorentz invariance. Type II DSM is proposed as a platform for topological superconductivity and Majorana fermions and is reported in $PtSe_2$ \cite{huang2016type}, $RbMgBi$ \cite{le2017three}, $VAl_3$ \cite{chang2017type}, and $YPd_2Sn$ \cite{guo2017type} classes. On the other hand, type III DSMs, also called critical DSMs, result from topologically protected tilted bands but have line-like Fermi surfaces with a topological invariant $N_2 =1$ and are theoretically predicted in $Zn_2 In_2 S_5$ \cite{huang2018black}, $Hf_x Zr_{1-x} Te_2$ \cite{fragkos2021type} and $Ni_3 In_2 X_2(X=S, Se)$ \cite{sims2023analogous}.

In order to induce band overlap within the bulk band structure, the emergence of topological behaviour typically requires a substantial perturbation. Spin-orbit coupling (SOC) commonly fulfills this criterion. Nevertheless, there exist alternative methods for achieving band crossings. Application of pressure can modify atomic orbital hybridizations or the strength of spin-orbit coupling, thereby leading to band inversion near the Fermi level and facilitating attractive topological phase transitions (TPTs) \cite{khan2023pressure}. The pressure-induced shift in bands may occur in either a uniaxial or triaxial manner, governed by the symmetry and structural constraints of the crystal. Several semimetals, including LaAs \cite{khalid2018topological}, LaSb \cite{zeng2016compensated}, TmSb \cite{wadhwa2019first}, TaAs \cite{zhou2016pressure} and YbAs \cite{singh2022pressure}, have been demonstrated to transition into topologically non-trivial phases under pressure. Furthermore, topological phases in materials like LaSb \cite{khalid2020trivial} and SnTe \cite{fragkos2019topological} have been observed when subjected to epitaxial strain, and these observations have been experimentally confirmed through angle-resolved photoemission spectroscopy (ARPES) in the case of SnTe \cite{fragkos2019topological}. 

In this paper, we have studied computationally predicted n-type thermoelectric material \cite{gorai2020computational} KCdP, which has a hexagonal crystal structure with a finite bulk band gap and act as a semiconductor. Applying suitable negative pressure allowed the band to shift and overlap between conduction and valence band along the $\Gamma$-A path, resulting in a topological phase transition from normal to triple point semimetal when SOC is absent, whereas when we apply SOC, the triple point converted into a DSM, which is confirmed by symmetry analysis and surface band calculation. Further increase in negative pressure introduced more bands to cross near the Fermi energy and converted to multiple Dirac points in KCdP. Analysis of the electronic band structure of KCdP reveals that the DSM phase in this material is induced by band inversion and is protected by crystal symmetry.

\section{Symmetry Analysis}\label{sec2}

In this section, we will explain the emergence of the exotic non-trivial topological behavior as a result of symmetry protection from the group theory approach. The compound KCdP  (space group 194 $P6_3/mmc$) possesses $D_{6h}$ point group.  The crossing between the conduction and valence bands occurs along the $\Gamma$–$A$ line (the $k_z$ axis) at 3\% negative triaxial pressure, where the little group is $C_{6v}$. The $C_{6v}$ little group contains the symmetry elements identity (E), sixfold ($C_6$), threefold ($C_3$), twofold ($C_2$) rotation about the $k_z$ axis and three $\sigma_v$, and three $\sigma_d$ mirror planes. The generators of this $C_{6v}$ little group are $C_{6z}$ and $\sigma_v$ (i.e, xz-mirror plane).

In the absence of SOC for $C_{6z}$ rotation operator, $C^6_{6z} = 1$ and $e^{i2\pi n} = 1$ so the eigenvalues of $C_{6z}$ operator are $e^{i2\pi n /6 }$, where n = 0,1,2,3,4,5 and lets say the corresponding eigenstates are $\psi_1$ , $\psi_2$ , $\psi_3$, $\psi_4$, $\psi_5$ and $\psi_6$, respectively. Under the action of $\sigma_v$ mirror plane $\psi_2$ changes to $\psi_6$ (i.e, $\tilde{\sigma_v}\psi_2 \longrightarrow \psi_6$) and $\psi_3$ changes to $\psi_5$ (i.e, $\tilde{\sigma_v}\psi_3 \longrightarrow \psi_5$ ), we have discussed about $\psi_1$ \& $\psi_4$ later in this section. It is important to notice here that, $C_{6z}$ doesn't commute with $\tilde{\sigma_v}$ (xz mirror plane). Two space group (symmetric) operators that anti-commute guarantee twofold degeneracy \cite{wieder2016spin}. The Hamiltonian (H) commutes with $\tilde{\sigma_v}$ i.e, $ [H,\tilde{\sigma_v}] = 0$. This commutation gives-

\begin{eqnarray}
    & H\tilde{\sigma_v}\psi_2 - \tilde{\sigma_v} H \psi_2 = 0\\ 
    & H \psi_6 - E_2 \tilde{\sigma_v}\psi_2 = 0\\
    & E_6 \psi_6 - E_2 \psi_6 = 0
\end{eqnarray}

In the above equation, it is clear that the energy eigenvalue, $E_2$ and $E_6$, corresponding to rotation eigenstates $\psi_2$ and $\psi_6$, respectively, are the same. Also, we can show that $E_3$ = $E_5$, where $E_3$ and $E_5$ are the energy eigenvalues corresponding to rotational eigenstates $\psi_3$ and $\psi_5$, respectively. Thus, the non-commutation of $\tilde{C_{6z}}$ and $\tilde{\sigma_v}$ makes sure that there must be two double degenerate bands that is spanned by \{$\psi_2,\psi_6$\} and \{$\psi_3,\psi_5$\} along the $k_z$ axis, which is invariant under both $\sigma_v$ and $C_{6z}$, which is clear from the band structure. The bands named as  $\Gamma_7$ and $\Gamma_8$, in the band structure along the $\Gamma - A$ line, correspond to $A_1$ and $E_2$ of $C_{6v}$ little group, which are one-dimensional  and two-dimensional bands, respectively. This confirms that the conduction band is non-degenerate, while the valence bands are doubly degenerate. Now speaking of $\psi_1$ and $\psi_4$, they can not be coupled together following the above Hamiltonian approach. So $\psi_1$ and $\psi_4$ remain non-degenerate. From the above analysis, we can conclude that the $C_{6v}$ symmetry group must contain both one and two-dimensional irreducible representations (IRs) when SOC is not included. Also, this can be verified from the character table of the $C_{6v}$ point group \cite{gao2021irvsp}.

 The reason why we claim $\psi_1$ and $\psi_4$ are non-degenerate is that these two states ($\psi_1$ \& $\psi_4$) are eigenstates of $C_{2z}$ operation and the $C_{2z}$ operation commutes with the $\sigma_d$ operation. This commutation implies that $\psi_1$ and $\psi_4$ are eigenstates of $\sigma_d$ operator too. Thus, this commutation implies that the transformation between $\psi_1$ and $\psi_4$ is not possible and hence they are non-degenerate under the act of $\sigma_d$ (yz-plane) mirror \cite{barman2020symmetry}. Now, if the non-degenerate conduction band intersects closely with the doubly degenerate valence band near the Fermi level, the emergence of a triply degenerate fermion becomes possible along the $\Gamma$-A line, as shown in supplementary Figure S1. Further increase in negative pressure to $5\%$, the conduction band would cross one additional doubly degenerate band to form double threefold fermions, as shown in supplementary Figure S3.

Now, in the presence of SOC, the eigenvalues for $C_{6z}$ rotation operator are $e^{i(n+1/2)2\pi/6}$ with n = 0 to 5, and we label the corresponding eigenstates in the same identical designation ($\psi_1$ to $\psi_6$) as discussed before for the without SOC case. With SOC in consideration, the act of $\sigma_v$ operation changes a little bit, i.e., $\sigma_v^{2} = -1$, in the presence of SOC, a full 2$\pi$ rotation induces an extra phase $e^{i\pi}$ for the half-integer spin. Now under the action of $\sigma_v$, the $\psi_2$ and $\psi_3$ transforms as: $\tilde{\sigma_v}\psi_2 \longrightarrow i\psi_6$ and $\tilde{\sigma_v}\psi_3 \longrightarrow i\psi_5$ as discussed earlier.  Due to the non-commutation of $C_{6z}$ and $\sigma_v$, the two double-degenerate bands spanned by  \{$\psi_2,\psi_6$\} and \{$\psi_3,\psi_5$\} are still enforced in the presence of SOC along the $k_z$ axis, which is invariant under both $\sigma_v$ and $C_{6z}$ operation.

Unlike without SOC case, here $\psi_1$ and $\psi_4$ make a degenerate eigen space in the presence of SOC. Since in earlier case we have seen that [$C_{2z}$, $\sigma_d$] = 0, we now define a new operator as the product of $\tilde{C_2}$ and $\tilde{\sigma_d}$ i.e., $\xi = \tilde{C_2} \tilde{\sigma_d}$. Now, $\xi^2 = 1$ due to the commutation of $\tilde{C_2}$ and $\tilde{\sigma_d}$  and their squares equal to -1 when SOC is considered. In the presence of time reversal symmetry (TRS), we can define a new operator, $\Theta=\xi T$. As $T^2 = -1$ in the presence of SOC, we have $\Theta^2 = \xi^2 T^{2} = -1$ along the $k_z$ axis, which is invariant under both $C_2$ and $\sigma_d$. Thus, along the $k_z$ axis, $\Theta^2 = -1$ acts as a local Kramers theorem and enforces degeneracy of $\psi_1$ and $\psi_4$ on the $C_{2z}$ axis. Therefore, $C_{6v}$ allows only two-dimensional IRs along the $C_6$ axis. Thus, any accidental band crossing on the $C_6$ axis will form a fourfold degenerate Dirac node. The Dirac nodes are protected from band hybridization-induced gap opening if the two doubly degenerate bands belong to different IRs of $C_{6v}$. The Kramers theorem states that Inversion (I) and TRS together enforce all bands to become doubly degenerate over the whole Brillouin zone, including $\Gamma$-A direction. However, according to the above analysis $C_{6v}$ subgroup alone give Dirac like band crossing.
 
\begin{figure*}[t]
\centering
\includegraphics[width=1\linewidth]{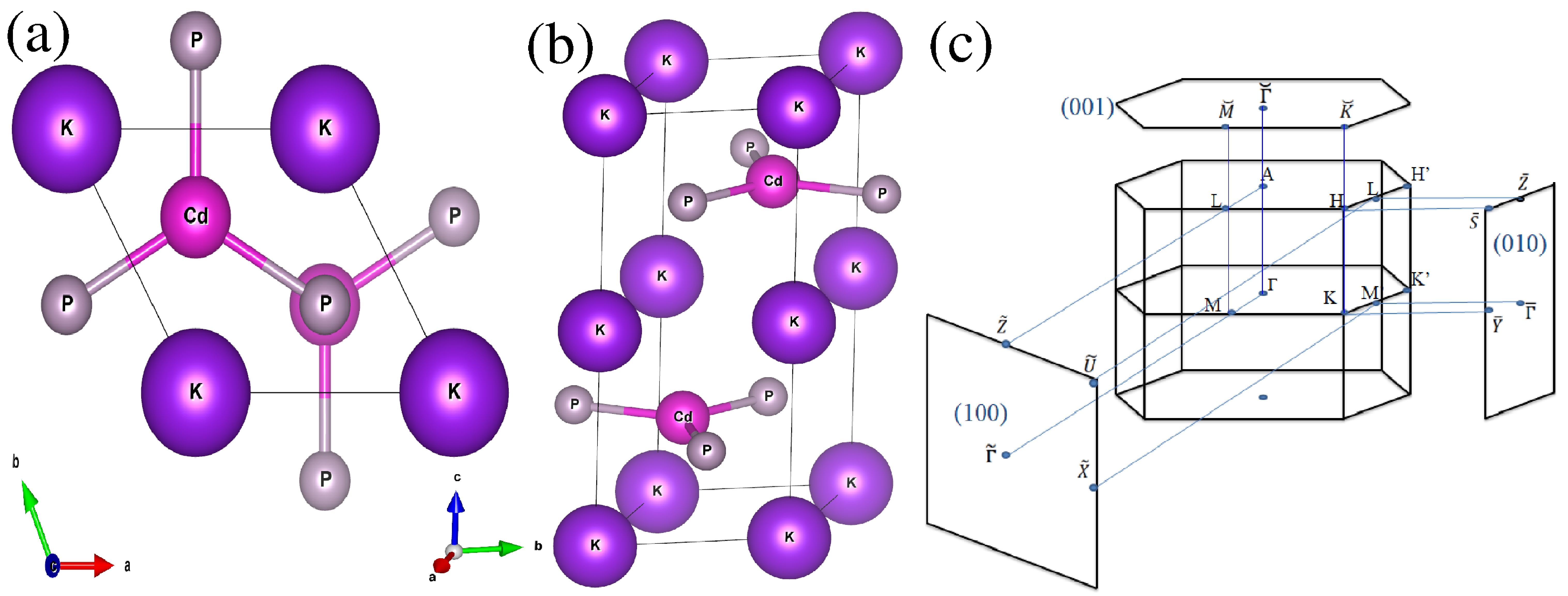}
\caption{\label{fig:epsart1} (a) Crystal structure of KCdP (b) Crystal structure of KCdP from different angle. (c) Bulk and surface Brillouin zone [(100),(010),(001)] of hexagonal crystal structure for KCdP.}
\end{figure*}

\begin{table}[!b]
\centering
\renewcommand{\arraystretch}{2} 
\caption{\label{demo-table} Optimized lattice parameters, pressure, and band gap of KCdP under ambient and negative triaxial pressure conditions.}
\begin{tabular}{||c|c|c|c|c|c||}
\hline\hline
S. No. & Pressure (\%) & Pressure (GPa) & Lattice Parameters & Band Gap (GGA-PBE) & Band Gap (HSE06) \\
\hline\hline
1 & 0  & 0.000  & a = 4.44 \AA, c = 10.18 \AA & 220 meV & 896 meV \\
\hline
2 & 3  & -2.558 & a = 4.57 \AA, c = 10.48 \AA & Dirac semimetal & 602 meV \\
\hline
3 & 5  & -3.550 & a = 4.65 \AA, c = 10.68 \AA & Dirac semimetal & 399 meV \\
\hline
4 & 7  & -4.228 & a = 4.75 \AA, c = 10.89 \AA & Dirac semimetal & 198 meV \\
\hline
5 & 10 & -4.888 & a = 4.88 \AA, c = 11.19 \AA & Dirac semimetal & Dirac semimetal \\
\hline\hline
\end{tabular}
\end{table}

\begin{figure*}[t]
\centering
\includegraphics[width=1\linewidth]{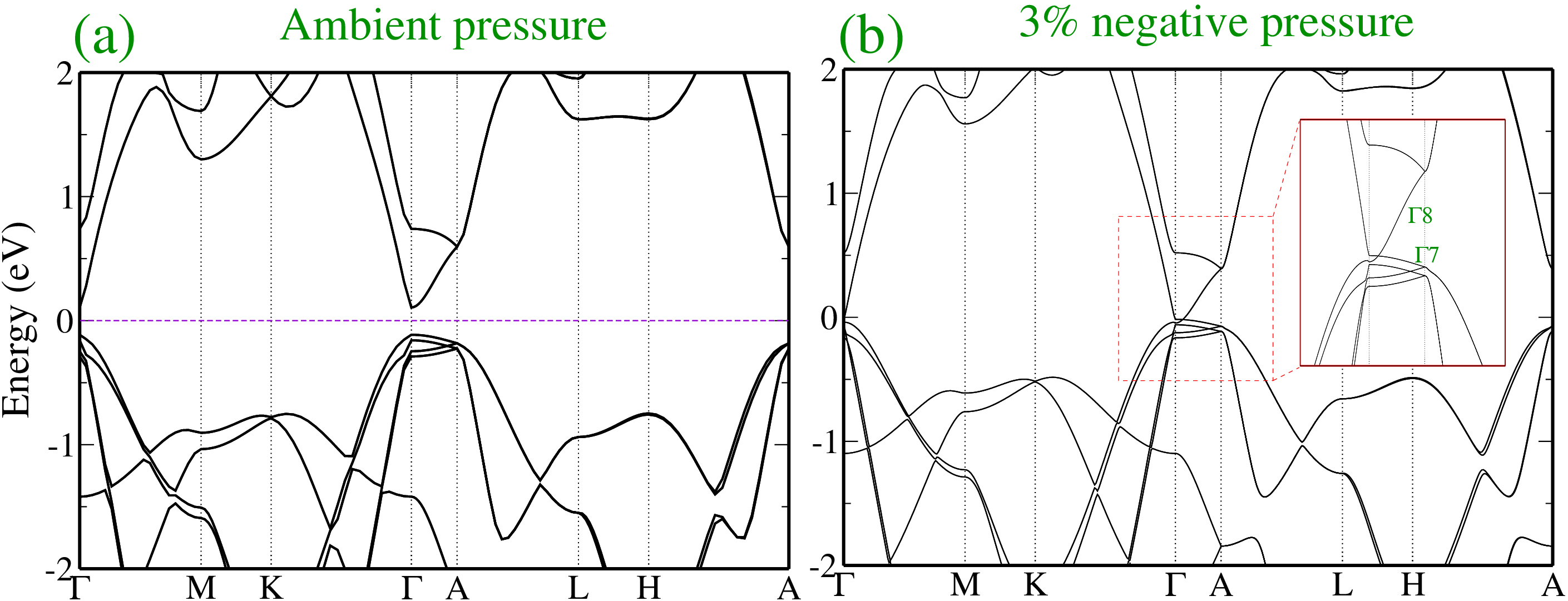}
\caption{\label{fig:epsart2} Electronic bulk band structure including the spin-orbit coupling of (a) KCdP with ambient pressure (b) KCdP with negative triaxial pressure of $3\%$.  }
\end{figure*}  

\begin{figure*}[t]
\includegraphics[width=1.0\linewidth]{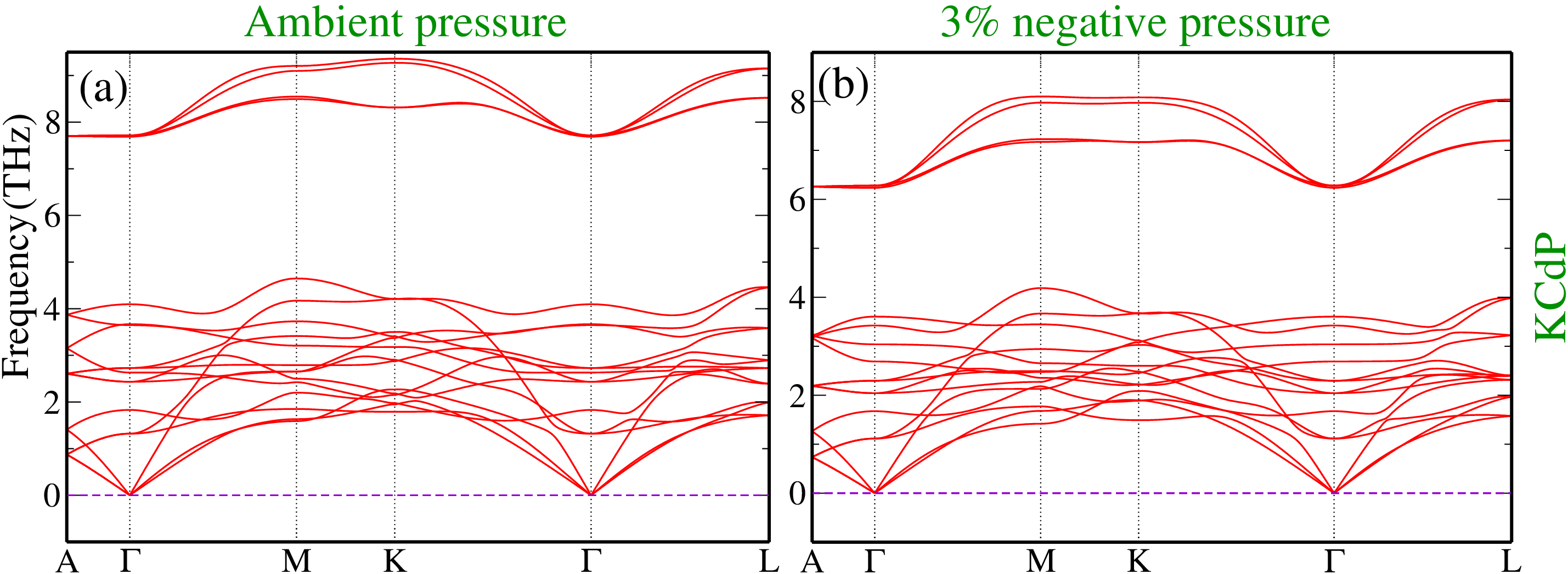}
\caption{\label{fig:epsart3} Phonon band structure of (a) KCdP with ambient pressure. (b) KCdP with negative triaxial pressure of $3\%$}
\end{figure*}

\begin{figure}[b!]
\includegraphics[width=1 \linewidth]{figure3.eps}
\caption{\label{fig:epsart4} (a) 3D Band inversion bulk band structure including the spin-orbit coupling. (b) Band induced Dirac semimetal: band inversion between $s$-orbital and $p_{x+y}$-orbital of P for KCdP.}
\end{figure} 
 
 \begin{figure*}[t]
 \centering
\includegraphics[width=1\linewidth]{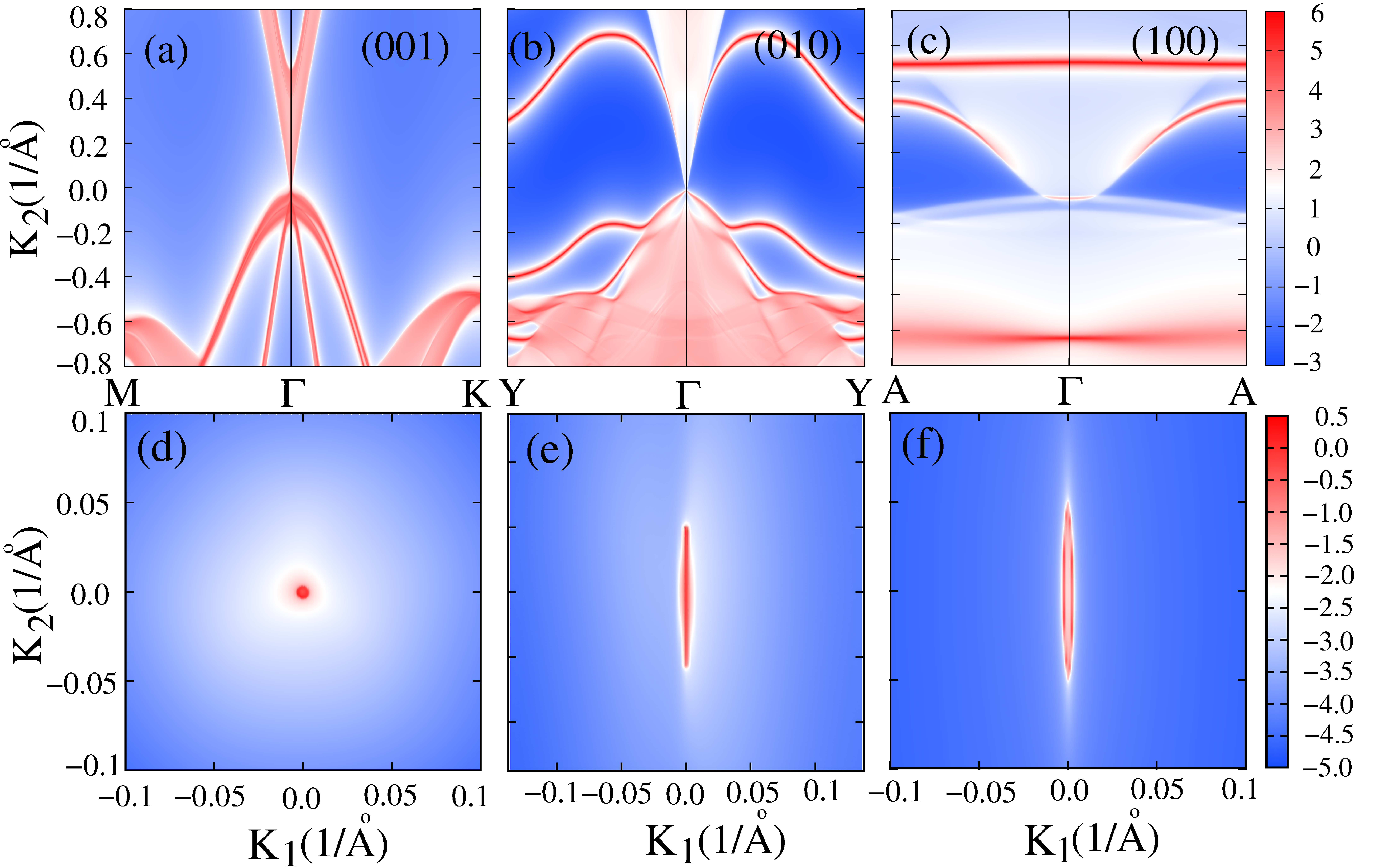}
\caption{\label{fig:epsart6} Surface state of KCdP: (a) (001) plane at M$-\Gamma-$K, (b) (010) plane at Y$-\Gamma-$Y and (c) (100) plane at A$-\Gamma-$A. Dirac point at Fermi surface corresponding to (d) (001) surface, (e) (010) surface and (f) (100) surface at the Fermi level ($E=E_f$).}
\end{figure*}

\section{Results and Discussion}\label{sec3}

First-principles calculation has been carried out for the previously reported n-type thermoelectric material KCdP \cite{gorai2020computational}. The three-dimensional material KCdP belongs to the space group $P6_3/mmc$ (194) and has a hexagonal crystal structure with optimized lattice parameters a=4.44 \AA, c=10.18 \AA, as shown in table-\ref{demo-table},  Figure-\ref{fig:epsart1} (a) and \ref{fig:epsart1} (b). The three-dimensional Brillouin zone and corresponding two-dimensional Brillouin zone for KCdP are shown in Figure-\ref{fig:epsart1} (c). In the search for topological material, the electronic properties of KCdP at ambient pressure have been calculated, and the band structure of KCdP considering the spin-orbital coupling has been shown in Figure-\ref{fig:epsart2} (a). In KCdP, the nature of the band is semiconducting, and one of the conduction bands shows a cone-like structure at the $\Gamma$-point. The KCdP material exhibits a larger direct band gap of 220 meV in the generalized gradient approximation (GGA)-Perdew–Burke–Ernzerhof (PBE) potential. The absence of imaginary phonon frequency, as evident from the phonon bandstructure of KCdP shown in Figure \ref{fig:epsart3} (a), ensures the dynamical stability of the material at ambient pressure.  

With the intent to observe the topological phase transition in KCdP, we applied negative triaxial pressure uniformly along all three crystallographic directions. As the lattice constant increased, the conduction and valence bands gradually approached each other. At a negative pressure of $3\%$, the conduction and valence bands overlapped at the Fermi level, indicating a topological phase transition in KCdP even in the absence of spin-orbit coupling (SOC), as shown in Supplementary Figure S1. It can be observed that the bands remain gapped along all paths of the Brillouin zone except the $\Gamma$--$A$ path in the $z$-direction. Band representation analysis further reveals that the valence band is doubly degenerate ($\Gamma_6$), whereas the conduction band is non-degenerate ($\Gamma_1$), resulting in the formation of a topological triple point semimetal as indicated in Supplementary Figure S1. 

When SOC is included in the $3\%$ negative triaxial pressure case, all the bands become doubly degenerate due to Kramers degeneracy. Consequently, the triply degenerate crossing transforms into a topological Dirac point at the Fermi level as shown in Figure~\ref{fig:epsart2} (b) and supplementary Figure S2. 

To verify the dynamical stability of the proposed crystal system in the negative triaxial pressure state, phonon dispersion calculations were performed. The phonon spectra of KCdP under negative triaxial pressure, as shown in Figure~\ref{fig:epsart3}(b), exhibit no imaginary phonon modes, confirming that the material remains stable beyond the transition state. These phonon studies further indicate the practical feasibility of synthesizing the pressure-induced material. For the present study, phonon dispersion curves were obtained both at ambient pressure and under negative triaxial pressure ($3\%$), with no imaginary frequencies observed in either case, confirming that KCdP is dynamically stable under the considered conditions.

Upon increasing the applied negative triaxial pressure to $5\%$, a further evolution in the band structure of KCdP is observed. In the absence of spin-orbit coupling (SOC), the previously identified non-degenerate conduction band crosses two doubly degenerate valence bands along the $\Gamma$-A direction. This results in the formation of two distinct topological triple points, highlighting the rich band topology of the system under negative triaxial pressure, as shown in supplementary Figure S3. These triple points emerge due to the symmetry-protected band crossings between bands of different IRs.

When SOC is introduced at the $5\%$ negative triaxial pressure state, the degeneracies of the bands are lifted and reorganized according to Kramers' theorem. The doubly degenerate conduction band crosses the four doubly degenerate valence bands, and as a consequence, triple points transform into four topological Dirac points are located near the Fermi level, as shown in supplementary Figure S4. These Dirac points are symmetry-protected and appear as fourfold degenerate crossings, arising from the interplay between time-reversal symmetry and crystal symmetry.  

The evolution from semiconductor to triple point semimetal (in the non‐SOC case) and semiconductor to Dirac semimetal (in the SOC case) under negative triaxial pressure confirms that KCdP undergoes a negative pressure‐driven topological phase transition. Such transitions demonstrate the tunability of the electronic structure with external perturbations and emphasize the potential of KCdP as a candidate material for exploring novel quasiparticles in condensed matter systems.

Furthermore, to provide an intuitive visualization of the Dirac cone, we present the three-dimensional (3D) band dispersion around the Dirac point. As shown in Figure~\ref{fig:epsart4}(a), two symmetry-protected Dirac points emerge within the first Brillouin zone due to a band crossing along the $\Gamma$–A high-symmetry line. Owing to time-reversal symmetry, this crossing appears symmetrically along both the negative $A$–$\Gamma$ and $\Gamma$–A directions. The 3D band dispersion exhibits a linear energy–momentum relationship along all relevant momentum directions, providing direct and unambiguous evidence of the Dirac semimetallic nature of KCdP under negative triaxial pressure. In the negative pressure phase, the semimetallic character is further confirmed by the linear crossing of doubly degenerate bulk bands, indicative of a topological Dirac semimetal (TDSM) phase arising from band inversion, similar to that observed in Na$_3$Bi and Cd$_3$As$_2$. In KCdP, this band inversion occurs between the $s$ and $p_{x+y}$ orbitals of phosphorus, as illustrated in Figure~\ref{fig:epsart4}(b), suggesting that experimental investigations could further validate the predicted TDSM phase.
 
Now, to confirm the topological properties of KCdP in negative triaxial pressure states, surface state calculation has been performed. KCdP has the topological surface state for (001), (010), and (100) at the $\Gamma$-point as shown in Figure-\ref{fig:epsart6}(a), \ref{fig:epsart6}(b), and \ref{fig:epsart6}(c), respectively, and the corresponding Fermi arc surface has been shown in Figure-\ref{fig:epsart6}(d), \ref{fig:epsart6}(e), and \ref{fig:epsart6}(f) at the Fermi level ($E=E_f$), respectively. The presence of surface states in the surface band structure and its Dirac points confirmed that the KCdP is a TDSMs in the negative triaxial pressure state. Further investigation shows that the DSM properties of this material is induced by band inversion between the s-orbital and $p_{x+y}$-orbital of Phosphorus, and the crossing of the bands indicates that this material belongs to type I DSMs similar to $Na_3Bi$ and $Cd_3As_2$ category. \\

In addition to the presence of symmetry-protected Dirac points, the existence of massless Dirac fermions with high Fermi velocity is a defining characteristic of Dirac materials, as exemplified by graphene. To further substantiate the Dirac fermionic nature of KCdP, we have calculated the Fermi velocity in the vicinity of the Dirac cone by extracting the slope of the linear band dispersion around the Dirac point, using
\begin{equation}
v_F = \frac{1}{\hbar}\left|\frac{dE}{dk}\right|.
\end{equation}
The estimated Fermi velocity is found to be of the order of $1.425 \times 10^{5}\,\mathrm{m\,s^{-1}}$. This value is comparable to those reported for other three-dimensional DSMs, although it is smaller than that of graphene ($\sim 10^{6}\,\mathrm{m\,s^{-1}}$) \cite{hwang2012fermi, elias2011dirac, lima2015controlling}. The reduced Fermi velocity is expected due to the three-dimensional nature of KCdP and the corresponding linear dispersion of the bands near the $\Gamma$ point. These results further confirm the presence of massless Dirac fermions in negative triaxial pressure state KCdP and reinforce its classification as a DSM.\\

\begin{figure}[h!]
\centering
\includegraphics[width=1\linewidth]{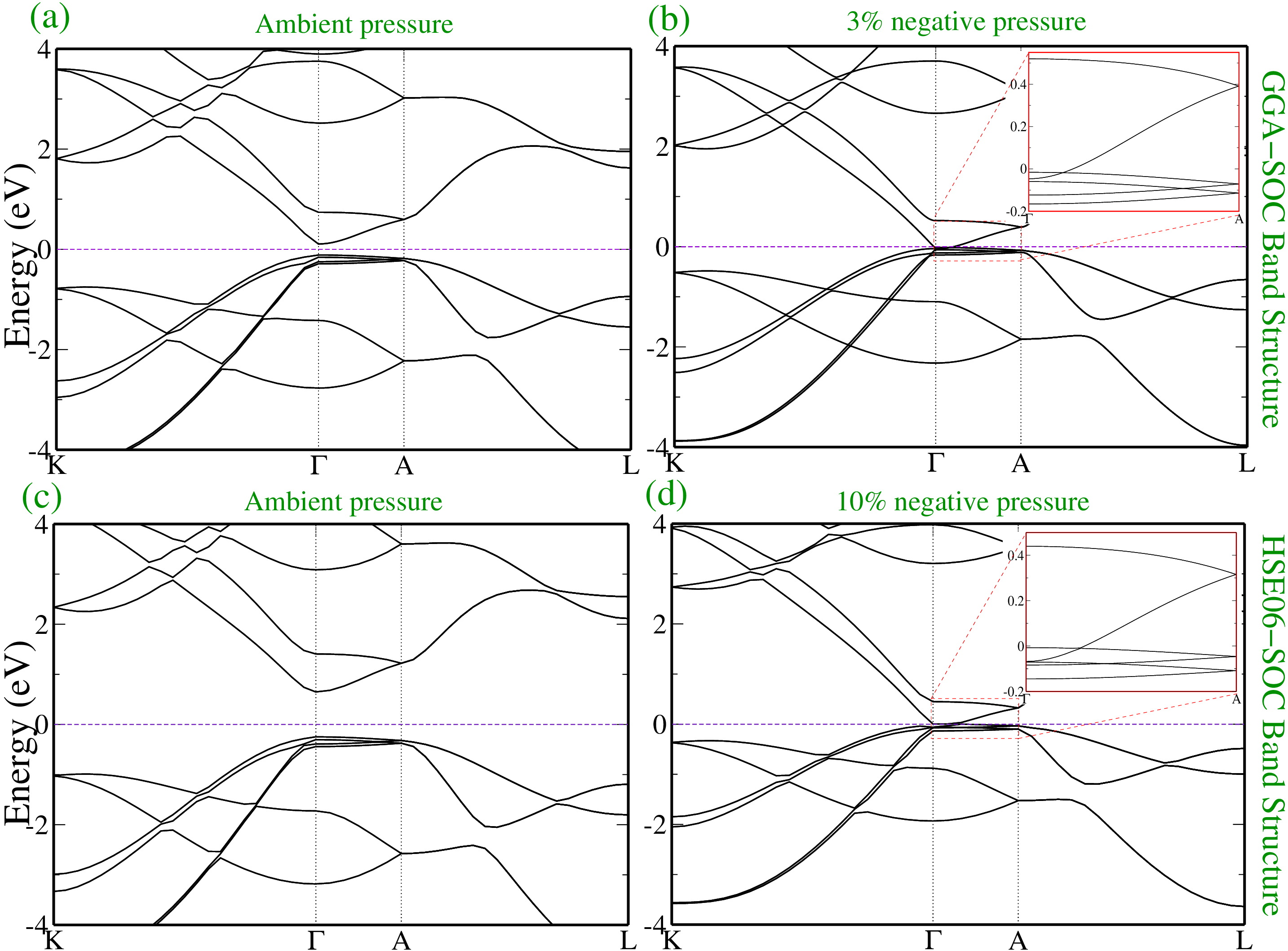}
\caption{\label{fig:epsart7} Electronic bulk band structures including spin–orbit coupling (SOC) calculated using the GGA functional: (a) KCdP at ambient pressure and (b) KCdP under negative triaxial pressure of 3\%. Corresponding results obtained using the HSE06 hybrid functional: (c) KCdP at ambient pressure and (d) KCdP under negative triaxial pressure of 10\%.}
\end{figure}

\textbf{\underline{Effect of HSE06 Hybrid Functional on the Electronic Band Structure}}\\

It is well known that the GGA exchange–correlation functional systematically underestimates the band gaps of semiconductors and insulators and may even misclassify semiconducting systems as metallic. In contrast, the hybrid HSE06 functional incorporates a fraction of exact exchange, leading to a more accurate description of electronic correlations and consequently providing band gaps that are in much better agreement with experimental observations \cite{heyd2003hybrid, yuan2018gga, xue2018improved, mao2022dft, yang2025fundamentals}.

\begin{figure*}[t!]
\centering
\includegraphics[width=0.6\linewidth]{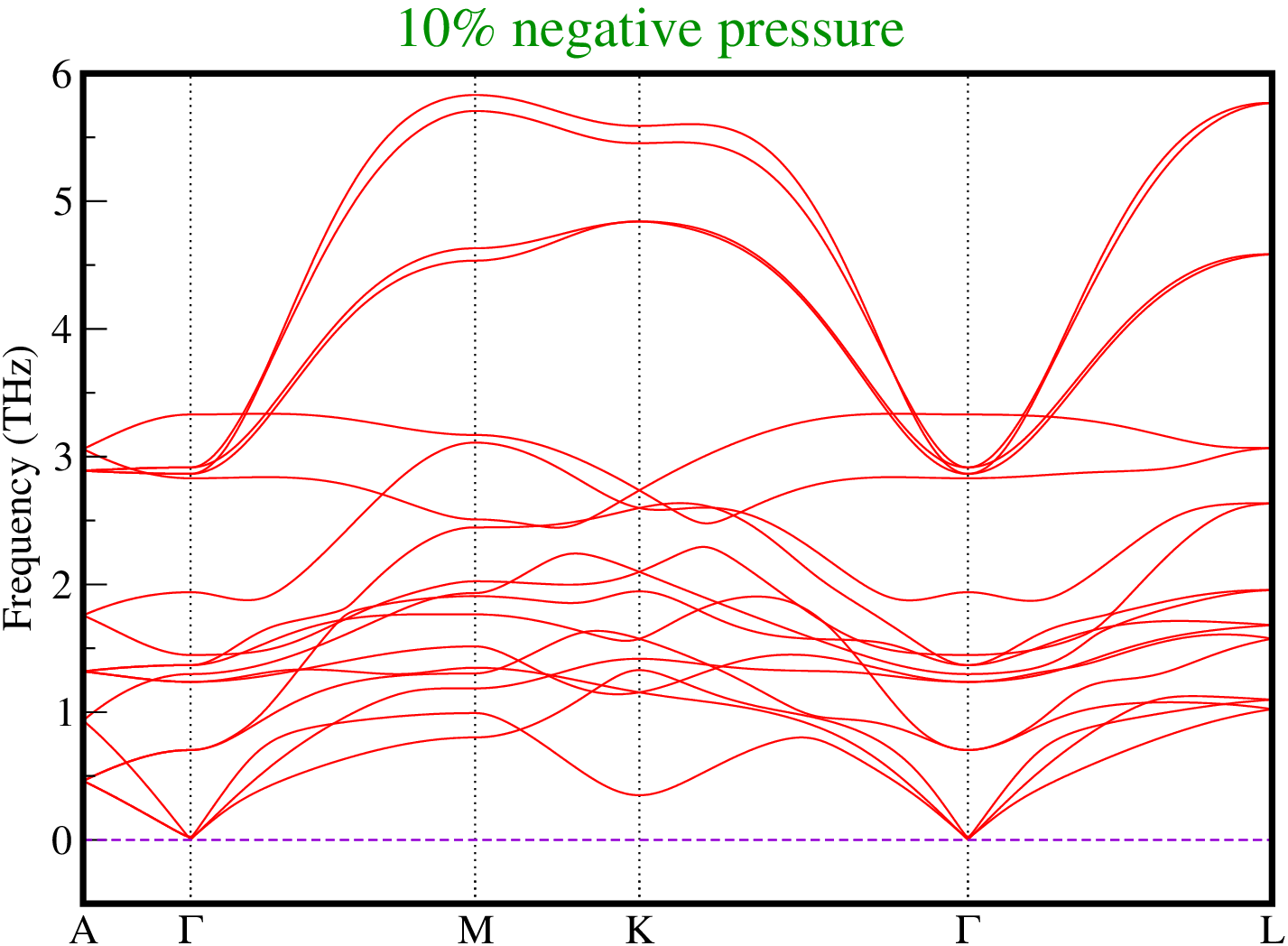}
\caption{\label{fig:epsart8} Phonon band structure of KCdP with 10\% of negative triaxial pressure}
\end{figure*}

For KCdP, despite the quantitative difference in the band gap values obtained from GGA-PBE and HSE06 as shown in table-\ref{demo-table}, the essential features of the electronic structure remain unchanged across the topological transition in both approaches. In particular, the band ordering and orbital character near the Fermi level at the $\Gamma$ high-symmetry point are preserved, demonstrating that the electronic band character is identical before and after the transition in both functionals as shown in Figure \ref{fig:epsart7}. This indicates that the difference between GGA-PBE and HSE06, is purely quantitative rather than qualitative. As a result, the critical negative triaxial pressure required to induce the topological phase transition shifts from approximately 3\% within GGA-PBE to about 10\% in the HSE06 calculations. Importantly, the nature of the topological phases on either side of the transition remains unchanged, with no additional band inversions or symmetry modifications introduced by the hybrid functional. Additionally, we have calculated the phonon band structure at 10\% negative triaxial pressure to demonstrate the dynamical stability of KCdP, as shown in Figure~\ref{fig:epsart8}. 

After the topological phase transition, the electronic band ordering in the vicinity of the Fermi level is found to be nearly identical when described using either the GGA-PBE or HSE06 exchange–correlation functionals. As a result, both approaches are expected to yield the same surface-state dispersion, and will not yield qualitatively new information \cite{weng2016topological, patel2024topological, sun2016pressure, mondal2019emergence}. On this basis, we adopt the computationally efficient GGA-PBE functional to construct the tight-binding Hamiltonian using maximally localized Wannier functions (MLWFs), which is then employed to confirm the TDSM phase in KCdP. Accordingly, the GGA-PBE surface-state calculations presented in the manuscript are sufficient to capture the essential topological physics of the system.

\section{Computational Methods}\label{sec4}

Electronic properties of the KCdP compound have been investigated by the first-principles calculation based on standard density functional theory \cite{kohn1965self} with the Projector Augmented-Wave (PAW) method with a plane-wave basis provided by the VASP package \cite{hafner2008ab, blochl1994projector, kresse1996efficient}. Generalized gradient approximation (GGA) with Projector augmented wave (PAW) \cite{torrent2010electronic} potentials have been utilized to incorporate the exchange-correlation function. Additionally, the calculations based on the hybrid HSE06 functional \cite{heyd2003hybrid} were carried out to refine the description of the electronic structure, as the GGA exchange-correlation functional is known to systematically underestimate the band gaps of semiconductors. The plane-wave basis energy cutoff of 500eV and $\Gamma$ -centered Monkhorst-pack \cite{monkhorst1976special} k-grid of $12\times 12\times 6$ for KCdP is used to perform self-consistent calculations (SCF). The structure, after the application of hydrostatic pressure, was relaxed by employing a conjugate-gradient scheme until the forces on each atom became less than 0.005 eV/Å. The band structure was then calculated including SOC, using the optimized structure. Maximally localized Wannier functions (MLWF) are used to develop the tight-binding model, which is used to calculate the surface states of the materials using the Wannier90 code \cite{mostofi2008wannier90}. WannierTools is used to obtain topological characteristics such as topological surface state and Fermi surface \cite{wu2018wanniertools}. A supercell with dimensions of $3\times 3\times 2$ has been constructed to calculate the phonon band structure for KCdP for ambient pressure $3\%$ and $10\%$ negative triaxial pressure. The phonon band structures were determined using Density Functional Perturbation Theory (DFPT), implemented in the phonopy code \cite{togo2015first,phonopy,phonopy1}.

\section{Conclusions}\label{sec5}
An n-type thermoelectric material, KCdP, has been investigated using first-principles calculations. At ambient conditions, KCdP exhibits a finite band gap and behaves as a normal semiconductor. Upon the application of negative triaxial pressure, the material undergoes a topological phase transition. In the absence of spin--orbit coupling (SOC), KCdP enters a triple-point semimetal phase, whereas inclusion of SOC drives the system into a Dirac semimetallic state. The Dirac semimetallic nature of KCdP under negative triaxial pressure is confirmed through symmetry analysis, surface band structure calculations, and the presence of Dirac points in the Fermi-arc surface states.
Importantly, the transition occurs within a minimal pressure range while preserving the dynamical stability of the crystal structure, making KCdP a promising candidate for experimental realization. Further analysis reveals that the DSM phase originates from a pressure-induced band inversion mechanism, similar to well-known DSMs such as $Na_3Bi$ and $Cd_3As_2$, and that KCdP belongs to the type-I DSM class under negative triaxial pressure.
Beyond the mere existence of Dirac points, the presence of massless Dirac fermions with high Fermi velocities near the Dirac cone is a defining characteristic of Dirac materials. Our results demonstrate that KCdP under negative triaxial pressure hosts such massless Dirac fermions, thereby reinforcing its classification as a DSM. Overall, our findings provide a comprehensive understanding of the pressure-driven topological properties of KCdP and open avenues for further experimental investigations and potential technological applications in the field of quantum materials.

\medskip
\textbf{Acknowledgements} \par 
This study received funding from the Anusandhan National Research Foundation (ANRF), Government of India (Grant No. CRG/2022/006419), and from the Council of Scientific and Industrial Research – Human Resource Development Group (CSIR-HRDG) under the ASPIRE scheme (Grant No. 03WS(006)/2023-24/EMR-II/ASPIRE). Additional support was provided by the Department of Science and Technology (DST), Government of India, through the DST-FIST grant (No. SR/FST/PSI/2017/5(C)) sanctioned to the Department of Physics, VNIT Nagpur. The authors also acknowledge the infrastructure and computational facilities made available by Visvesvaraya National Institute of Technology (VNIT), Nagpur. Author P.S. is grateful for the high-performance computing assistance from the National Param Supercomputing Facility (NPSF) at C-DAC Pune. Author S.K.G. thanks the Ministry of Human Resource Development (MHRD), Government of India, for financial assistance through an institute fellowship.

\medskip

%
\bibliographystyle{MSP}
\bibliography{biblo1.bib}


\end{document}



\title{Supplementary Information for ``Pressure-Induced Topological Dirac Semimetallic Phase in KCdP"}

\author{Shivendra Kumar Gupta}
\email{shivendrakumarg900@gmail.com}
\affiliation{Department of Physics, Visvesvaraya National Institute of Technology, Nagpur, 440010, India \\}
\author{Nikhilesh Singh}
\author{Saurabh Kumar Sen}%
 \author{Nagarjuna Patra}%
\author{Poorva Singh}
\email{poorvasingh@phy.vnit.ac.in}
\affiliation{Department of Physics, Visvesvaraya National Institute of Technology, Nagpur, 440010, India \\}

\maketitle

\onecolumngrid




\section{Electronic Band Structure of \NoCaseChange{KCdP} at 3\% negative triaxial pressure without and with SOC}

\begin{figure}[!h]
\includegraphics[width=0.78 \linewidth]{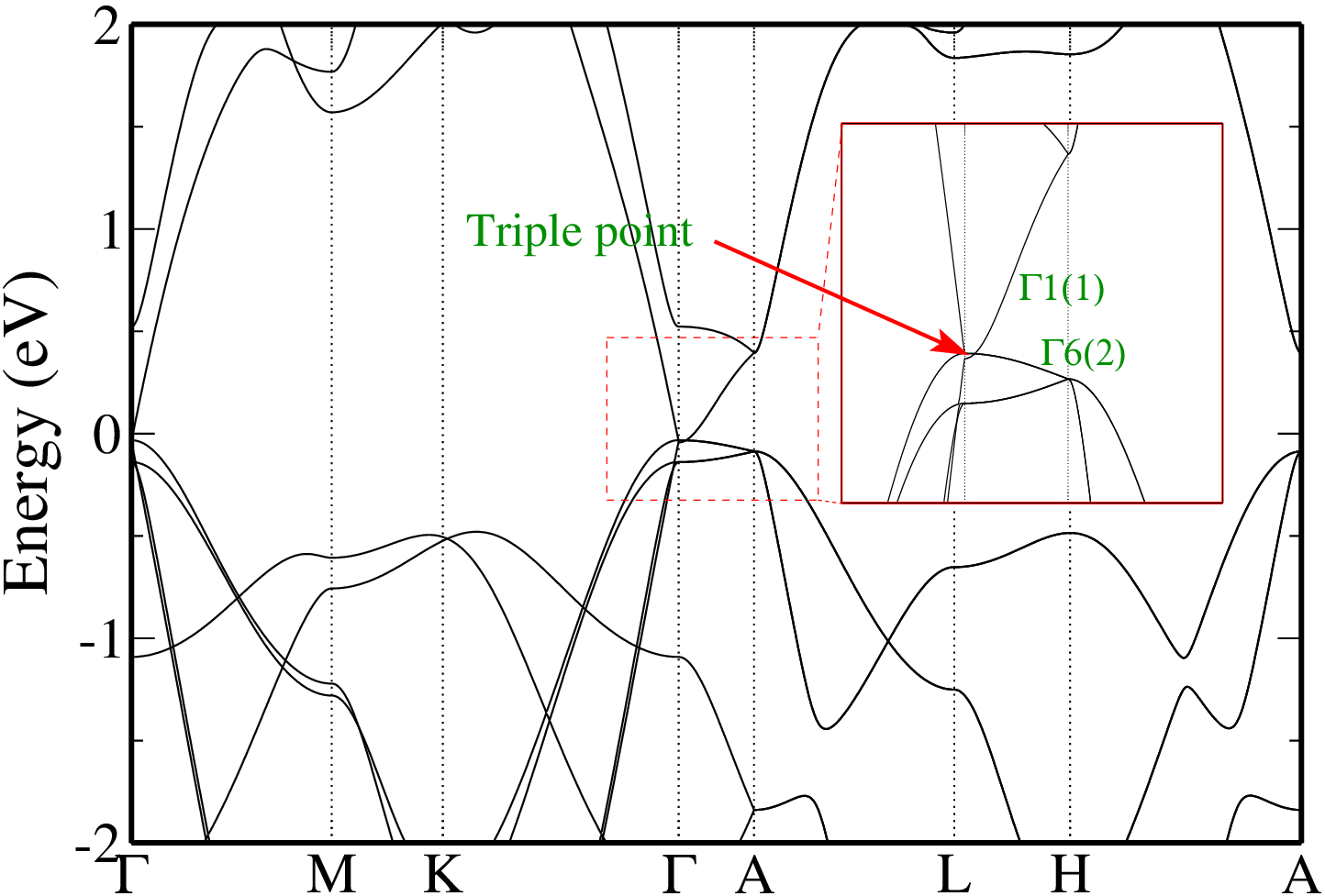}
 \caption{\label{fig:figure6} Electronic Band Structure of KCdP at 3\% negative triaxial pressure without SOC.}

 \end{figure}

 \begin{figure}[!h]
\includegraphics[width=.78 \linewidth]{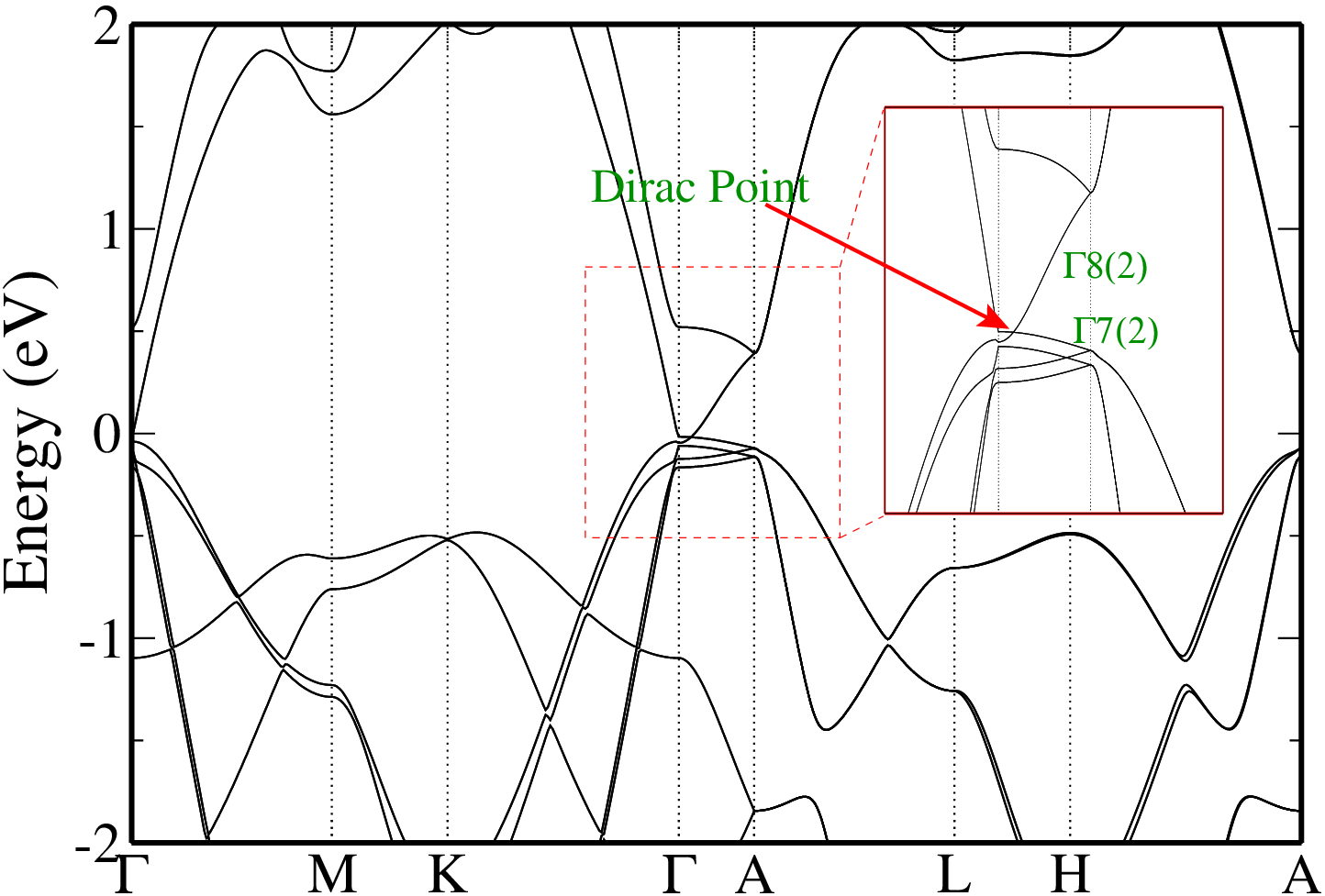}
 \caption{\label{fig:figure6} Electronic Band Structure of KCdP at 3\% negative triaxial pressure with SOC.}
 \end{figure}
 \pagebreak
 \section{Electronic Band Structure of \NoCaseChange{KCdP} at 5\% negative triaxial pressure without and With SOC}
 
\begin{figure}[!h]
\includegraphics[width=0.78 \linewidth]{BAND_5_wsoc.eps}
 \caption{\label{fig:figure6} Electronic Band Structure of KCdP at 5\% negative triaxial pressure without SOC.}

 \end{figure}
 
 \begin{figure}[!h]
\includegraphics[width=0.78 \linewidth]{BAND_5_soc.eps}
 \caption{\label{fig:figure6} Electronic Band Structure of KCdP at 5\% negative triaxial pressure with SOC.}

 \end{figure}

\pagebreak 
\bibliography{bib}